\documentclass[twocolumn,showpacs,preprintnumbers,amsmath,amssymb,prc,nofootinbib]{revtex4}

\usepackage{graphicx}% Include figure files
\usepackage{dcolumn}% Align table columns on decimal point
\usepackage{bm}% bold math

\begin{document}

\preprint{APS/123-QED}

\title{Proton induced reaction cross section measurements on Se isotopes for the astrophysical
$p$ process}

\author{Gy.~Gy\"urky}%
\email{gyurky@atomki.hu}
\author{Zs.~F\"ul\"op}
\author{E.~Somorjai}

\affiliation{%
Institute of Nuclear Research (ATOMKI), H-4001 Debrecen, Hungary
}%

\author{M.~Kokkoris}
\author{S.~Galanopoulos}
\author{P.~Demetriou}
\author{S.~Harissopulos}
\affiliation{Institute of Nuclear Physics, NCSR ``Demokritos'',
153.10 Aghia Paraskevi,
Athens, Greece
}%

\author{T.~Rauscher}
\affiliation{Departement f\"ur Physik und Astronomie, Universit\"at
Basel, Basel, Switzerland}

\author{S.~Goriely}
\affiliation{Institut d' Astronomie et d' Astrophysique, Universit\'e 
Libre de Bruxelles, Brussels, Belgium}
\date{\today}% It is always \today, today,

\begin{abstract}
As a continuation of a systematic study of reactions relevant to
the astrophysical $p$ process, the cross sections
of the $^{74,76}$Se(p,$\gamma$)$^{75,77}$Br and
$^{82}$Se(p,n)$^{82}$Br reactions
have been measured at energies from
1.3 to 3.6~MeV using an activation
technique. The results are compared to the
predictions of Hauser-Feshbach statistical model calculations
using the NON-SMOKER and MOST codes.
The sensitivity of the calculations to variations in the optical proton
potential and the nuclear level density was studied. Good agreement between
theoretical and experimental reaction rates was found for the reactions
$^{74}$Se(p,$\gamma$)$^{75}$Br and $^{82}$Se(p,n)$^{82}$Br.
\end{abstract}



\maketitle

\section{Introduction}

Details of the nature of the astrophysical $p$ process \cite{woosley90,rayet95} 
producing the rare, proton-rich, stable isotopes of elements beyond Ni
still remain elusive. 
It has been shown that such proton-rich isotopes can be produced by
photodisintegrations in massive stars \cite{rau02,arn03}, 
involving ($\gamma$,n),
($\gamma$,p), and ($\gamma$,$\alpha$) reactions at stellar temperatures
exceeding 10$^9$ K. Depending on intricate details of the stellar
structure and evolution, $p$ nuclides are mostly produced in the final
explosion of a massive star ($M > 8 M_\odot$) as a core-collapse
supernova or in pre-explosive oxygen burning episodes \cite{rau02,arn03}.
Despite the fact that $p$ nuclei can be produced consistently with
solar ratios over a wide range of nuclei in such scenarios, there remain
deficiencies in a few regions, for mass numbers $A<124$ and $168\leq
A\leq 200$. The
problem is especially pronounced in the Mo-Ru region where the
$p$ isotopes are strongly underproduced. 
This fact motivates the search for alternative
or additional ways to produce these nuclides. 
Proton capture and photodisintegration processes in exploding
carbon-oxygen white dwarfs have been suggested as a source \cite{how91,jose02},
or thermonuclear explosions in the proton-rich layer accreted on the
surface of a neutron star in a binary system with mass flow from the
main-sequence companion star \cite{schatz,schatz01}.

Common to all approaches is that the modelling requires
a large network of hundreds of nuclear reactions involving stable 
nuclei as well as unstable, proton-rich nuclides.
It is well known that ($\gamma$,$\alpha$) reactions are important at
higher masses, whereas ($\gamma$,p) or proton capture is important for
the production of the less massive $p$ nuclei.
The relevant astrophysical reaction rates calculated from the
cross sections are inputs to this network, therefore
their knowledge is essential for $p$ process calculations.
While there are compilations of neutron capture data along the
line of stability, there are still
very few charged-particle cross sections determined experimentally
\cite{ful96,sau97,som98,bork98,chlo99,haris01,gyurky01,ozkan02,galanopulos03},
despite big experimental efforts in recent years. Thus, the
$p$ process rates involving charged projectiles
are still based mainly on (largely untested) theoretical cross
sections obtained from Hauser-Feshbach statistical model
calculations.

The aim of our systematic study is to contribute to the existing
database of measured cross sections
relevant to the astrophysical $p$ process, and to check the reliability of
the statistical model calculations over an extensive set of nuclides.
This way the uncertainties in the $p$ process abundance calculations arising
from nuclear physics input can be constrained.

In this paper we present measurements of the $^{74}$Se(p,$\gamma$)$^{75}$Br,
$^{76}$Se(p,$\gamma$)$^{77}$Br, and $^{82}$Se(p,n)$^{82}$Br reactions
in the astrophysically
relevant energy range using an activation technique. The two proton
capture reactions and their inverses 
are directly playing a role in the synthesis of
$^{74}$Se, whereas the (p,n) reaction can be used as a further test of
statistical model calculations. The choice of these
reactions is further elaborated in Sec.\ \ref{sec:reacs} and the
experimental method is described in Sec.\ \ref{sec:expmethod}.
The resulting cross sections and astrophysical $S$ factors
(given in Sec.\ \ref{sec:expres}) are
 compared with results obtained with the two Hauser-Feshbach
statistical model codes NON-SMOKER \cite{nonsmoker,nonsmoker0,nonsmoker1} and MOST
\cite{most} in Sec.\ \ref{sec:comp}.
Both codes are based on the statistical theory of Hauser and Feshbach \cite{hau}
but use different models for the nuclear ingredients of the
calculations. By using both codes we are able to investigate the effects
of a broader range of nuclear level densities and optical model
potentials. The astrophysical reaction rates derived from our new
experimental data are given in the concluding Sec.\ \ref{sec:concl}.

\section{Investigated reactions}
\label{sec:reacs}

The element Se has six stable isotopes with mass numbers $A=74$,
76, 77, 78, 80 and 82 having isotopic abundances of 0.89\%,
9.36\%, 7.63\%, 23.78\%, 49.61\% and 8.73\%, respectively.
Proton capture reactions on these isotopes lead to Br isotopes
among which $^{79}$Br and $^{81}$Br are stable. Therefore the cross
sections of the $^{78,80}$Se(p,$\gamma$)$^{79,81}$Br reactions
cannot be measured using an activation technique. The half life of
$^{78}$Br, i.e.\ the reaction product of $^{77}$Se(p,$\gamma$)$^{78}$Br,
is too short (T$_{1/2}$~=~6.49~min.) for our experimental method
(see experimental details). Thus the aim of the measurement was to
determine the proton capture cross section of three reactions:
$^{74,76,82}$Se(p,$\gamma$)$^{75,77,83}$Br. However, in the case of
$^{82}$Se the
(p,n) channel opens already at $E_{\rm lab}$~=~891~keV and, therefore, competes
strongly with the (p,$\gamma$) reaction in the whole energy range
investigated in the present work. Moreover, the strongest
$\gamma$ transition following the $\beta$ decay of $^{83}$Br 
has a very low relative intensity 
(only 1.2\,\% of the decays lead to the emission of this 529.6\,keV
$\gamma$-ray). 
The latter two facts made it impossible to observe the decay of $^{83}$Br 
and to measure 
the $^{82}$Se(p,$\gamma$)$^{83}$Br cross section. However, 
the $^{82}$Se(p,n)$^{82}$Br cross section could be determined.

\begin{figure}
\resizebox{\columnwidth}{!}{\rotatebox{270}{\includegraphics{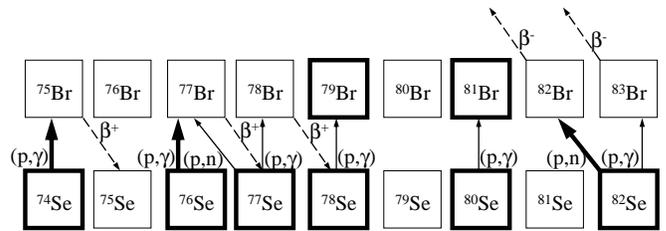}}}
\caption{The relevant part of the chart of nuclides with the 
decay of the reaction products. The stable isotopes are represented 
by bold squares.}
\label{nuclide}
\end{figure}

In summary, the cross sections of three reactions have been measured: 
$^{74,76}$Se(p,$\gamma$)$^{75,77}$Br, and $^{82}$Se(p,n)$^{82}$Br.
The astrophysically relevant energy range (Gamow--window) for these 
reactions in the temperature range from T\,=\,1.8$\times10^9$K to 3.3$\times10^9$K 
spans from 1.25 to 3.87\,MeV. This energy region was covered by the experiment.
The relevant part of the chart of nuclides can be seen in
Fig.\ \ref{nuclide} where the proton induced reactions and the 
decay of the reaction products can
also be seen. The decay parameters used for the analysis are
summarized in Table \ref{decay}.

\begin{table}
\caption{Decay parameters of the Br product nuclei taken from
literature.}
\begin{ruledtabular}
\begin{tabular}{ccccc}
\parbox[t]{1.5cm}{\centering{Product \\ nucleus}} &
\parbox[t]{1.5cm}{\centering{Half life [hour]}} &
\parbox[t]{1.5cm}{\centering{Gamma \\energy [keV]}} &
\parbox[t]{1.5cm}{\centering{Relative \\ intensity \\ per decay [\%]}} &
\parbox[t]{1.5cm}{\centering{Reference}} \\
\hline $^{75}$Br &  1.612 $\pm$ 0.002 & 286.6 & 88 $\pm$ 5 &
\cite{farhan90}
\\
             &                 &  141.3 & 6.6 $\pm$ 0.5 & \\
&&&&\\ $^{77}$Br & 57.036 $\pm$ 0.006 &  239.0 & 23.1 $\pm$ 0.5  &
\cite{farhan97} \\
             &                 & 520.6  & 22.4 $\pm$ 0.6 &  \\
             &                 & 297.2  & 4.16 $\pm$ 0.21 &  \\
&&&&\\
$^{82}$Br & 35.3 $\pm$ 0.02 & 776.5  & 83.5 $\pm$ 0.8 & \cite{king95} \\
          &                 & 554.3  & 70.8 $\pm$ 0.7 &  \\
          &                 & 619.1  & 43.4 $\pm$ 0.4 & \\

\end{tabular} \label{decay}
\end{ruledtabular}
\end{table}

\section{Experimental procedure}
\label{sec:expmethod}
\subsection{Target properties}

The targets were made by evaporating metallic Se with natural isotopic 
abundance onto 
a thick Al backing. Natural targets have the advantage that the 
three investigated reactions can be studied simultaneously in a single activation 
procedure. Aluminum is ideal as backing material because no long-lived activity 
is produced during its bombardment with protons
in the investigated energy range. Moreover,
Al can easily be distinguished from the Se in the RBS spectrum
(RBS was used to monitor the target stability, see below).

The target thickness was measured with the
proton induced X-ray emission (PIXE) technique 
at the PIXE set-up of the ATOMKI \cite{PIXE}. According to PIXE results,
the target thickness was ranging from 200 to
700\,$\mu$g/cm$^2$, corresponding to a proton energy loss of 10\,keV (at 3.6\,MeV) to 60\,keV (at 1.3\,MeV), 
respectively. The thicker targets (500-700\,$\mu$g/cm$^2$)
were used at the lower part of the bombarding energy range, where the low 
cross section results in very low induced $\gamma$-activity. The proton energy loss has been
calculated with the SRIM code \cite{SRIM}.

\subsection{Activation}

\begin{figure}
\resizebox{\columnwidth}{!}{\rotatebox{270}{\includegraphics{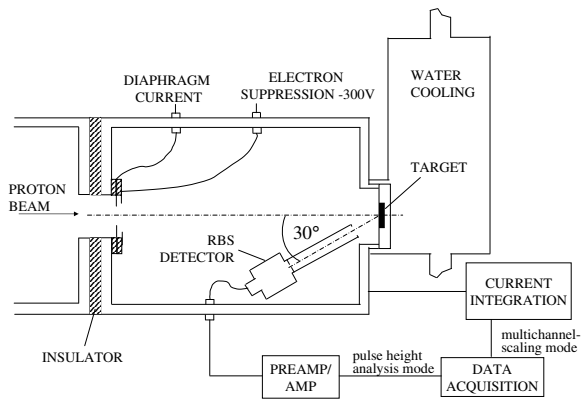}}}
\caption{Schematic view of the target chamber.}
\label{chamber}
\end{figure}

The activations were carried out at the 5~MV Van de Graaff
accelerator of the ATOMKI by irradiating the Se targets with a proton beam. 
The energy range from E$_p$=1.3 to 3.6~MeV was covered with 100 -- 300\,keV steps.
The schematic view of the target
chamber can be seen in Fig.\ \ref{chamber}. After the last beam
defining aperture the whole chamber served as a Faraday-cup to
collect the accumulated charge. A secondary electron suppression
voltage of $-300$ V was applied at the entrance of the chamber. Each
irradiation lasted about 10 hours with a beam current of
typically 5 to 10 $\mu$A. Thus, the collected charge varied between
180 and 360 mC. The current was kept as stable as possible but to
follow the changes the current integrator counts were recorded in
multichannel scaling mode, stepping the channel in every minute. This recorded 
current integrator spectrum was then used for the analysis solving the differential 
equation of the population and decay of the reaction products numerically 
(see eqs. 5-10 in ref \cite{sau97}).

A surface barrier detector was built into the chamber at
$\Theta$=150$^\circ$ relative to the beam direction to detect the
backscattered protons and to monitor the target stability this way.
The RBS spectra were taken continuously and stored regularly
during the irradiation. In those cases when target deterioration
was found, the irradiation was repeated with another target. 
Final cross section results were derived only from those measurements 
where no noticable target deterioration was found, i. e. the growth of 
the Se peak area in the RBS spectrum was uniform with collected charge 
within the statistical uncertainty (below 1\%).

The beam was wobbled across the last diaphragm to have a uniformly
irradiated spot of diameter of 8~mm on the target. The target
backing was directly water cooled with an isolated water
circulating system.

Between the irradiation and $\gamma$-counting, a waiting time of 30 minutes 
was inserted in order to let the disturbing short lived
activities decay.

\subsection{Detection of induced $\gamma$-radiation}

The $\gamma$ radiation following the $\beta$-decay of the produced
Br isotopes was measured with a HPGe detector of 40\% relative
efficiency. The target was mounted in a holder at a distance of
10\,cm from the 
end of the detector cap. The whole system was
shielded by 10~cm thick lead against laboratory background.

The $\gamma$ spectra were taken for at least 12 hours and stored
regularly in order to follow the decay of the different reaction
products.

The absolute efficiency of the detector was measured with
calibrated $^{133}$Ba, $^{60}$Co and $^{152}$Eu sources in the 
same geometry used for the measurement. Effect of the finite sample size 
(beam spot of 8 mm in diameter) was measured by moving the point-like calibration 
sources over this surface and measuring the difference in efficiency. This was then 
included in the 7\% error of detector efficiency.

\begin{figure}
\resizebox{\columnwidth}{!}{\rotatebox{270}{\includegraphics{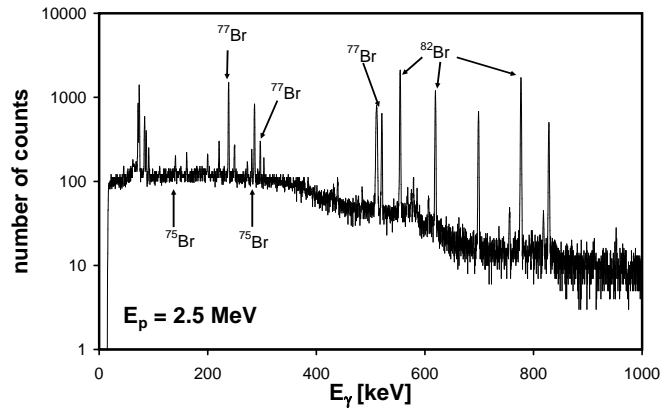}}}
\caption{Activation $\gamma$ spectrum after irradiating a target with 2.5\,MeV protons. 
The $\gamma$ lines used for the analysis are indicated by arrows. The not labeled peaks
are either from laboratory background or beam-induced background on impurities in the target and 
backing or $\gamma$ lines from Se + p reaction products which were not used for the analysis.}
\label{gammaspec}
\end{figure}

Fig.\ \ref{gammaspec} shows an off-line $\gamma$-spectrum taken
after irradiation with 2.5~MeV protons in the first 1h counting interval. 
The $\gamma$ lines used for the analysis are indicated by arrows.

Taking into account the detector efficiency and the relative intensity of the 
emitted gamma-rays, coincidence summing effects were for all three reactions 
below 1\% and were neglected.

\section{Experimental results}
\label{sec:expres}

Tables \ref{74tab} -- \ref{82tab} summarize the experimental results for the three 
studied reactions. The quoted errors of the center-of-mass energies given
in the first column correspond to the energy loss in the targets calculated with the SRIM code. 
The error of the cross section ($S$ factor) values 
is the quadratic sum of the following partial errors: efficiency of the HPGe detector ($\sim$7\%), 
number of target atoms ($\sim$6\%), current measurement (3\%), uncertainty of the level parameters 
found in literature ($\leq$6\%), counting statistics (0.1 to 15\%).

At $E_{\rm p}$\,=\,2.176\,MeV proton bombarding 
energy the $^{77}$Se(p,n)$^{77}$Br reaction 
channel opens, having the same final nucleus as $^{76}$Se(p,$\gamma$)$^{77}$Br.
Thus, above this energy these two reactions cannot be distinguished with the
activation technique 
using targets with natural isotopic abundance. The measured cross section 
and S--factor values 
above the threshold are actually the weighted sum of the two cross sections: 
$\sigma_\mathrm{exp}$\,=\,$\sigma_1$\,+\,0.82$\cdot$$\sigma_2$, where $\sigma_1$ and 
$\sigma_2$ are the cross sections of $^{76}$Se(p,$\gamma$)$^{77}$Br and $^{77}$Se(p,n)$^{77}$Br, 
respectively and 0.82 stands for the isotopic ratio of $^{77}$Se and $^{76}$Se: 7.63\%/9.36\%.

\begin{table}
\caption{Experimental cross section and $S$ factor of the
$^{74}$Se(p,$\gamma$)$^{75}$Br reaction}

\begin{ruledtabular}
\begin{tabular}{r@{\hspace{0.2cm}$\pm$\hspace{-0.7cm}}lr@{\hspace{0.2cm}$\pm$\hspace{-0.7cm}}
lr@{\hspace{0.2cm}$\pm$\hspace{-0.7cm}}l} \multicolumn{2}{c}{\hspace{-0.2cm}E$_{\rm c.m.}$} &
\multicolumn{2}{c}{\hspace{-0.4cm}Cross section} &
 \multicolumn{2}{c}{\hspace{-0.5cm}$S$ factor} \\
\multicolumn{2}{c}{\hspace{-0.2cm}[keV]} & \multicolumn{2}{c}{\hspace{-0.4cm}[$\mu$barn]} &
\multicolumn{2}{c}{\hspace{-0.5cm}[10$^6$ MeV$\cdot$b]} \\
\hline
1455 & 27 & 3.6 & 0.7 & 5.6 & 1.1 \\
1569 & 12 & 10.8 & 1.9 & 6.6 & 1.2 \\
1658 & 22 & 20.8 & 3.2 & 6.4 & 1.0 \\
1766 & 12 & 41.7 & 6.7 & 6.1 & 1.0 \\
1858 & 20 & 60.7 & 9.0 & 5.0 & 0.8 \\
1954 & 22 & 94.0 & 14 & 4.4 & 0.7 \\
2057 & 18 & 119 & 18 & 3.2 & 0.5 \\
2159 & 14 & 191 & 28 & 3.1 & 0.5 \\
2310 & 11 & 366 & 55 & 3.0 & 0.5 \\
2456 & 13 & 467 & 69 & 2.1 & 0.3 \\
2752 & 13 & 1039 & 154 & 1.6 & 0.2 \\
3049 & 12 & 1561 & 231 & 0.98 & 0.15 \\
3348 & 8 & 1888 & 282 & 0.54 & 0.08 \\
3547 & 6 & 2966 & 483 & 0.53 & 0.09
\end{tabular} \label{74tab}
\end{ruledtabular}

\end{table}

\begin{table}
\caption{Experimental cross section and $S$ factor of the
$^{76}$Se(p,$\gamma$)$^{77}$Br reaction. Note that above 2.148 MeV
c.m.\ energy the $^{77}$Se(p,n)$^{77}$Br reaction has a
contribution to the quoted experimental values. See text for details.}

\begin{ruledtabular}
\begin{tabular}{r@{\hspace{0.2cm}$\pm$\hspace{-0.7cm}}lr@{\hspace{0.2cm}$\pm$\hspace{-0.7cm}}
lr@{\hspace{0.2cm}$\pm$\hspace{-0.7cm}}l} \multicolumn{2}{c}{\hspace{-0.2cm}E$_{\rm c.m.}$} &
\multicolumn{2}{c}{\hspace{-0.4cm}Cross section} &
 \multicolumn{2}{c}{\hspace{-0.5cm}$S$ factor} \\
\multicolumn{2}{c}{\hspace{-0.2cm}[keV]} & \multicolumn{2}{c}{\hspace{-0.4cm}[$\mu$barn]} &
\multicolumn{2}{c}{\hspace{-0.5cm}[10$^6$ MeV$\cdot$b]} \\
\hline
1456 & 27 & 9.5 & 0.8 & 14.8 & 1.3 \\
1569 & 12 & 10.4 & 0.9 & 6.3 & 0.6 \\
1658 & 22 & 21.8 & 2.2 & 6.8 & 0.7 \\
1767 & 12 & 59.6 & 6.2 & 8.8 & 0.9 \\
1859 & 20 & 53.8 & 5.3 & 4.4 & 0.4 \\
1954 & 22 & 93.1 & 9.2 & 4.4 & 0.4 \\
2058 & 18 & 147 & 14 & 4.0 & 0.4 \\
2160 & 14 & 207 & 20 & 3.4 & 0.3 \\
2311 & 11 & 796 & 78 & 6.5 & 0.7 \\
2457 & 13 & 1182 & 116 & 5.3 & 0.5 \\
2753 & 13 & 3200 & 312 & 4.9 & 0.5 \\
3050 & 12 & 6467 & 631 & 4.0 & 0.4 \\
3349 & 8 & 11529 & 1143 & 3.3 & 0.3 \\
3548 & 6 & 17252 & 2044 & 3.1 & 0.4
\end{tabular} \label{76tab}
\end{ruledtabular}

\end{table}

\begin{table}
\caption{Experimental cross section and $S$ factor
of the $^{82}$Se(p,n)$^{82}$Br reaction}

\begin{ruledtabular}
\begin{tabular}{r@{\hspace{0.2cm}$\pm$\hspace{-0.7cm}}lr@{\hspace{0.2cm}$\pm$\hspace{-0.7cm}}
lr@{\hspace{0.2cm}$\pm$\hspace{-0.7cm}}l} \multicolumn{2}{c}{\hspace{-0.2cm}E$_{\rm c.m.}$} &
\multicolumn{2}{c}{\hspace{-0.4cm}Cross section} &
 \multicolumn{2}{c}{\hspace{-0.5cm}$S$ factor} \\
\multicolumn{2}{c}{\hspace{-0.2cm}[keV]} & \multicolumn{2}{c}{\hspace{-0.4cm}[$\mu$barn]} &
\multicolumn{2}{c}{\hspace{-0.5cm}[10$^6$ MeV$\cdot$b]} \\
\hline
1258 & 31 & 0.75 & 0.14 & 8.2 & 1.5 \\
1457 & 27 & 3.6 & 0.3 & 5.6 & 0.5 \\
1571 & 12 & 11.8 & 1.2 & 7.2 & 0.7 \\
1660 & 22 & 17.6 & 1.7 & 5.5 & 0.5 \\
1768 & 12 & 44.1 & 4.3 & 6.5 & 0.6 \\
1860 & 20 & 79.8 & 7.6 & 6.6 & 0.6 \\
1956 & 22 & 137 & 13 & 6.5 & 0.6 \\
2060 & 18 & 219 & 21 & 5.9 & 0.6 \\
2162 & 14 & 367 & 35 & 6.0 & 0.6 \\
2313 & 11 & 742 & 71 & 6.1 & 0.6 \\
2459 & 13 & 1142 & 109 & 5.1 & 0.5 \\
2755 & 13 & 3330 & 317 & 5.1 & 0.5 \\
3052 & 12 & 6778 & 647 & 4.2 & 0.4 \\
3352 & 8 & 12448 & 1208 & 3.6 & 0.4 \\
3552 & 6 & 19329 & 2257 & 3.5 & 0.4
\end{tabular} \label{82tab}
\end{ruledtabular}

\end{table}

\section{COMPARISON WITH PREDICTIONS}
\label{sec:comp}

In Figs.\ \ref{fig:74}--\ref{fig:82} we show the
comparison of
the experimental data with theoretical results.
At first, we discuss the standard predictions of the
Hauser-Feshbach codes MOST and NON-SMOKER, shown in the part (a) of each 
figure. Two different combinations of
nuclear level densities (NLD) and optical potentials (OMP) are used by
each code, the OMP of \cite{bau} with the NLD of
\cite{dem} in MOST, and the OMP of
\cite{jeu} with the NLD of \cite{rau} in NON-SMOKER.
From the figures it is apparent that for the Se isotopes
studied herein, the results of the former combination
are systematically higher (by a factor of 2) than the latter
at all studied energies.
Concerning the energy dependence, a decrease at low energy is seen for
$^{82}$Se(p,n)$^{82}$Br which seems not to be present in the data.
The predictions of the latter input
combination are in very good agreement with the data for energies
larger than 1.7 MeV in
$^{74}$Se(p,$\gamma$)$^{75}$Br, but cannot reproduce the observed
energy dependence at smaller energies.
They are higher than experiment by factors 1.25--2
for $^{76}$Se(p,$\gamma$)$^{77}$Br above 1.8 MeV and cannot reproduce
the experimental variation in the $S$ factor at lower energies.
They are able to reproduce
the data very well for the $^{82}$Se(p,n)$^{82}$Br reaction above 1.7
MeV. Again, the features in the experimental $S$ factors below that
energy cannot be reproduced. However, the deviations do not exceed a factor of
1.5. As will become evident in Sec.\ \ref{sec:concl}, these deviations at the
lower end of the measured energy range do not contribute significantly
to the astrophysical reaction rate for the temperatures given there.

In summary, contrary to some previous measurements of (p,$\gamma$)
reactions (see e.g.\ \cite{gyurky01}),
the standard predictions using the nuclear inputs of \cite{jeu,rau} are 
in good agreement with the present experimental
data and most of other proton capture data\footnote{Note: Ref.\ \cite{bork98} 
points out that the theoretical
values in \cite{sau97} were mistakenly given too
high. Using the correct value, there is good agreement also with the
experiment of \cite{sau97}.}. Even more so, if one considers that the
calculations use global parameters which are not fine-tuned to the
specific nuclei involved. For such global calculations, a 30\% deviation
averaged over all nuclei
is not unusual, similarly to what was found for neutron capture
\cite{rau}. Locally, larger errors are possible, of course.
Within this range of uncertainty the predictions agree very
well, especially regarding the astrophysical rates as given in Sec.\
\ref{sec:concl}. However, as can already be seen from the comparison of the
standard predictions, the results are sensitive to the nuclear inputs,
even though the reaction mechanism, and thus the reaction model to be
applied, is unambiguous. Therefore,
to better understand the contributions of different
model inputs it is necessary to disentangle the effects of mainly the
NLD and the OMP. This will also help to constrain future, improved
parametrizations of these nuclear properties.
For this purpose, additional calculations were
performed using both codes and varying the NLD and OMP models. 
The combination of NLDs and OMPs used in each calculation are summarized
in Table \ref{tab:hfinputs}. In the following sections, we discuss each
reaction separately.
\begin{table*}
\caption{\label{tab:hfinputs}Overview of the inputs in the
Hauser-Feshbach calculations.}
\begin{ruledtabular}
\begin{tabular}{lll}
\multicolumn{1}{c}{Label}&\multicolumn{1}{c}{Level
density}&\multicolumn{1}{c}{Optical potential}\\
\hline
INP-1 \footnotemark[1] \footnotemark[2]\footnotetext[1]{Calculated with
MOST} \footnotetext[2]{Standard prediction as available at
\url{http://www-astro.ulb.ac.be/Html/hfr.html} (version 09/12/2002)}&
Demetriou \& Goriely (2001) \protect\cite{dem}&Bauge et al.\ (2001)
\protect\cite{bau}\\
INP-2 \footnotemark[1]&Demetriou \& Goriely (2001) \protect\cite{dem}&
Jeukenne et al.\ (1977)
\protect\cite{jeu}\\
INP-3 \footnotemark[1]&Demetriou \& Goriely (2001) \protect\cite{dem}&
Koning (2002)
\protect\cite{kon}\\
INP-4 \footnotemark[1]&Thielemann et al.\ (1986) \protect\cite{thi}&
Jeukenne et al.\
(1977) \protect\cite{jeu}\\
\hline
INP-5 \footnotemark[3] \footnotemark[4] \footnotetext[3]{Calculated with
NON-SMOKER} \footnotetext[4]{Standard prediction as published in
\protect\cite{nonsmoker0,nonsmoker1}; also available
at \url{http://nucastro.org/reaclib.html}}&Rauscher et al.\
(1997) \protect\cite{rau}&Jeukenne et al.\
(1977) \protect\cite{jeu}\\
INP-6 \footnotemark[3]&Rauscher et al.\ (1997) \protect\cite{rau}&
equivalent square well
\protect\cite{holm}\\
INP-7 \footnotemark[3]&Rauscher et al.\ (1997) \protect\cite{rau}&
Becchetti \& Greenlees
(1969) \protect\cite{bec}\\
INP-8 \footnotemark[3]&Rauscher et al.\ (1997) \protect\cite{rau}&Perey (1963)
\protect\cite{per}\\
INP-9 \footnotemark[3]&Holmes et al.\ (1976) \protect\cite{holm}&
Jeukenne et al.\ (1977)
\protect\cite{jeu}\\
INP-10 \footnotemark[3]&Thielemann et al.\ (1986) \protect\cite{thi}&
Jeukenne et al.\
(1977) \protect\cite{jeu}
\end{tabular}
\end{ruledtabular}
\end{table*}

\subsection{$^{74}$Se(p,$\gamma$)$^{75}$Br}
\label{sec:74}

\begin{figure}
\resizebox{\columnwidth}{!}{\rotatebox{270}{\includegraphics{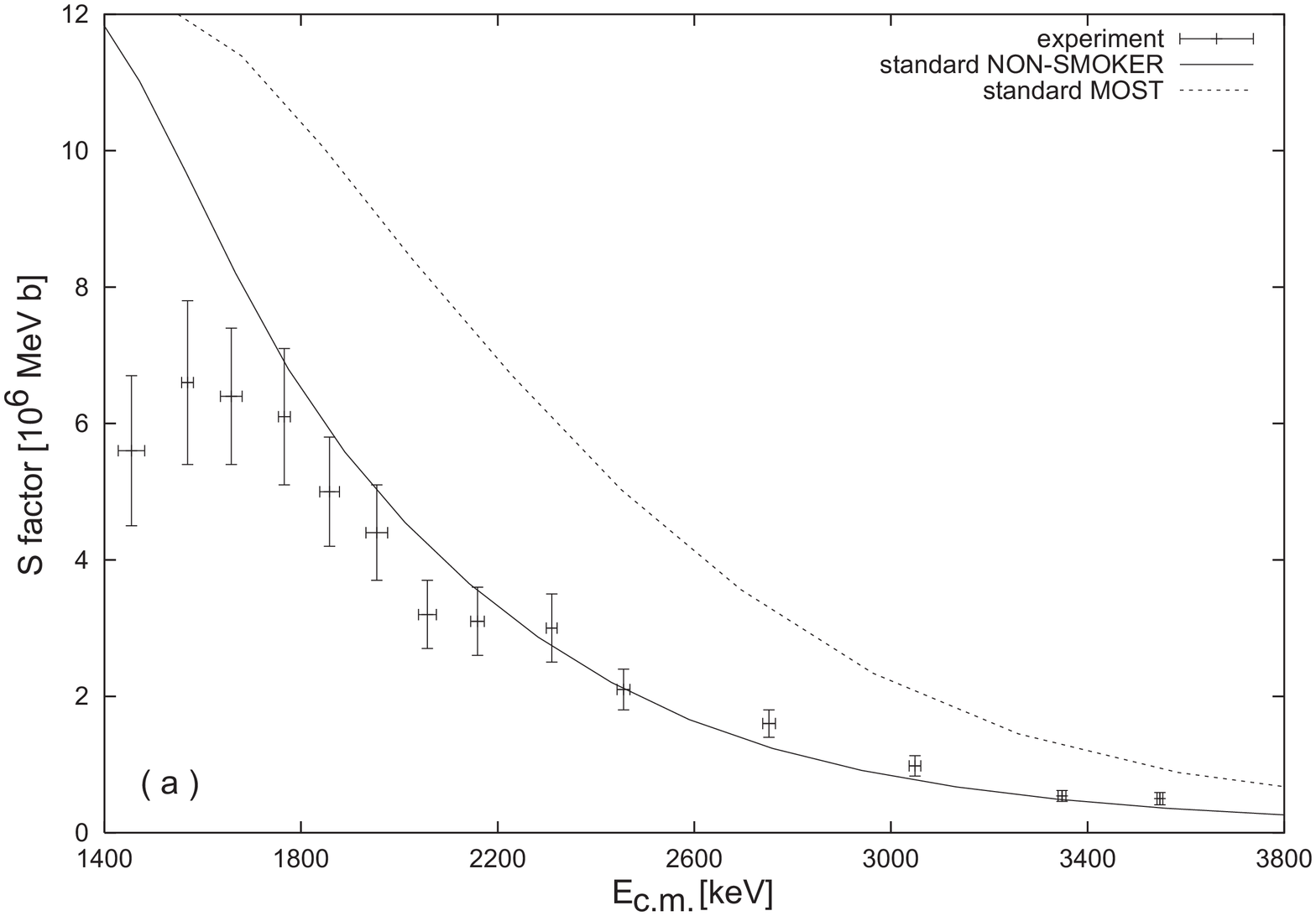}}}
\resizebox{\columnwidth}{!}{\rotatebox{270}{\includegraphics{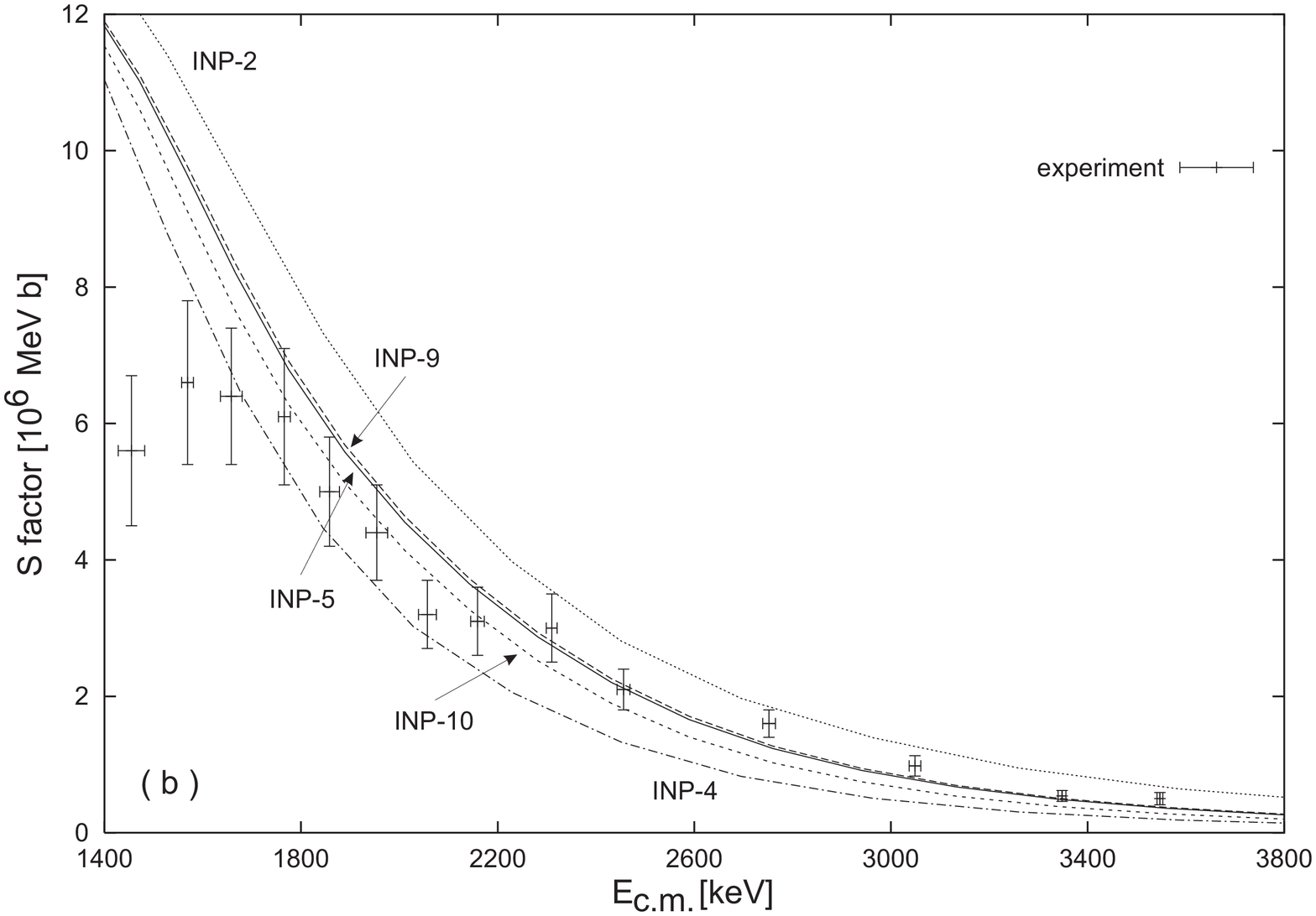}}}
\resizebox{\columnwidth}{!}{\rotatebox{270}{\includegraphics{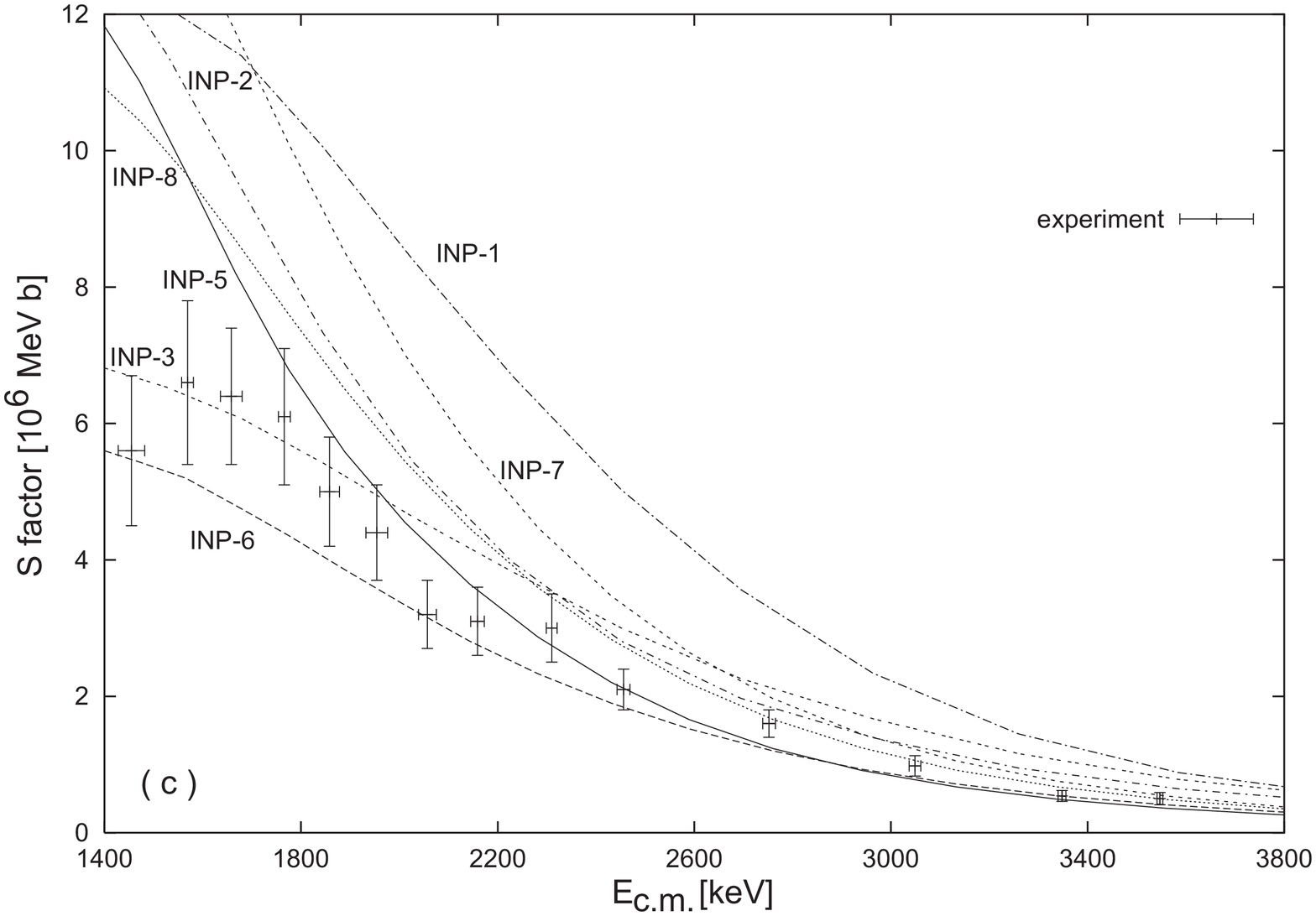}}}
\caption{Experimental $S$ factor of the reaction $^{74}$Se(p,$\gamma$)$^{75}$Br
and statistical model calculations. The standard predictions of MOST
(INP-1) and NON-SMOKER (INP-5) are compared to the data (a). Below, 
the effect of varying the nuclear level density
with fixed proton optical potential \protect\cite{jeu} is shown (b), 
as well as the effect of varying the optical potential (c). 
In the part (c) of the
figure, the curves labeled INP-1 to INP-3 include the level density of
\protect\cite{dem}, the ones labeled INP-5 to INP-8 the one of
\protect\cite{rau}. See Table \protect\ref{tab:hfinputs} for an explanation
of the labels.}
\label{fig:74}
\end{figure}

The comparison of the results for different NLDs and OMPs is shown in
the parts (b) and (c) of Fig.\ \ref{fig:74}. The labels are
explained in Table \ref{tab:hfinputs}.
It should be noted that INP-1 and INP-5 are the default predictions also
shown on part (a) in Fig.\ \ref{fig:74}. Moreover, INP-4 and INP-10 both use the same
level density {\em and} optical potential. The remaining difference
between them has to be attributed to
further differences in other nuclear inputs (such as the photon width)
because the basic approach
is the same in both calculations.
One has to keep in
mind this additional small difference when comparing results obtained
from the two codes (results INP-1 to INP-4 and INP-5 to INP-10, respectively).

As can be seen from the figure, the dependence on the OMP
is much stronger than the sensitivity to the NLD,
especially at the lower end of the studied energy range. All of the
calculations with the same OMP but different NLDs
show a similar energy dependence
whereas a change in the potential not only leads to different
absolute values but also to different excitation functions. 
At the low energies studied here, with no other open channels than
(p,$\gamma$), the cross section depends only on the incident proton
transmission coefficients, the NLD in the compound nucleus
determining the photon widths, and the NLD in the target entering the
compound-elastic proton channel. The effect of the NLD in the latter channel
is small, as transitions with higher proton energies to low-lying states
will be dominating and because these states are experimentally and
explicitly included in the calculation.
From Fig. \ref{fig:74} part (c) it is evident that none of the OMPs
are able to reproduce the data at low
energies, except for the ones of \cite{kon} (INP-3) and \cite{holm} (INP-6). 
The equivalent square-well potential of
\cite{holm} and the phenomenological OMP of \cite{kon} are able to give
the most satisfactory overall description of the data. The latter
phenomenological potential has been obtained for energies from 1 keV up
to 200 MeV,  using a unique, flat functional form for the energy
dependence. This may be the reason why it gives a relatively flatter
variation with energy compared to the other OMPs. What is somewhat
surprising, is the good agreement obtained with the simple equivalent
square-well potential.

The OMPs with the largest deviation are the ones of \cite{bec} (INP-7) and
\cite{bau} (INP-1).
The former OMP has been derived for energies above 20 MeV, so
it is not surprising that it fails. The latter is based on microscopic
nuclear matter calculations to start with, but has been subsequently
re-adjusted to reproduce an extensive database of reaction observables.
However, the reliability of the resulting renormalization coefficients
has only been tested for energies in the 10--30 MeV range. The
overprediction of the data could thus be attributed to inappropriate
values of the renormalization coefficients at low energies.
Another important effect might be caused by deformation which we further discuss
in Sec.\ \ref{sec:concl}.

\subsection{$^{76}$Se(p,$\gamma$)$^{77}$Br}
\label{sec:76}

In this case, the data cannot be straightforwardly compared with the
predictions because
of the complication with the $^{77}$Se(p,n) channel. As discussed in
Sec.\ \ref{sec:expres},
the experiment was not able to distinguish between $^{76}$Se(p,$\gamma$)
and $^{77}$Se(p,n) because they have the same final nucleus $^{77}$Br.
Therefore, only the points measured below the (p,n) threshold at
$E_{\rm p}$\,=\,2.176 MeV, are included in the
comparison.

\begin{figure}
\resizebox{\columnwidth}{!}{\rotatebox{270}{\includegraphics{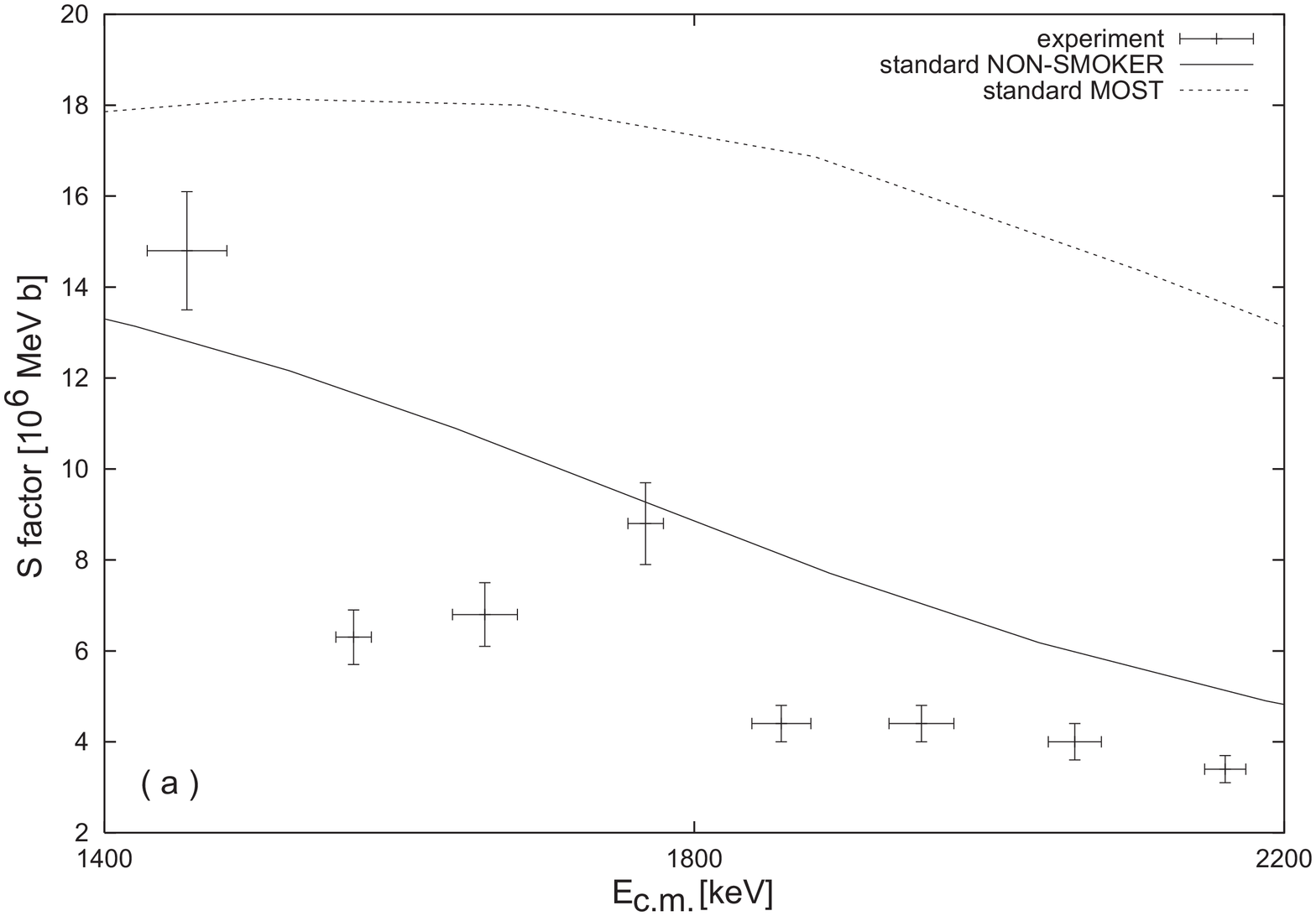}}}
\resizebox{\columnwidth}{!}{\rotatebox{270}{\includegraphics{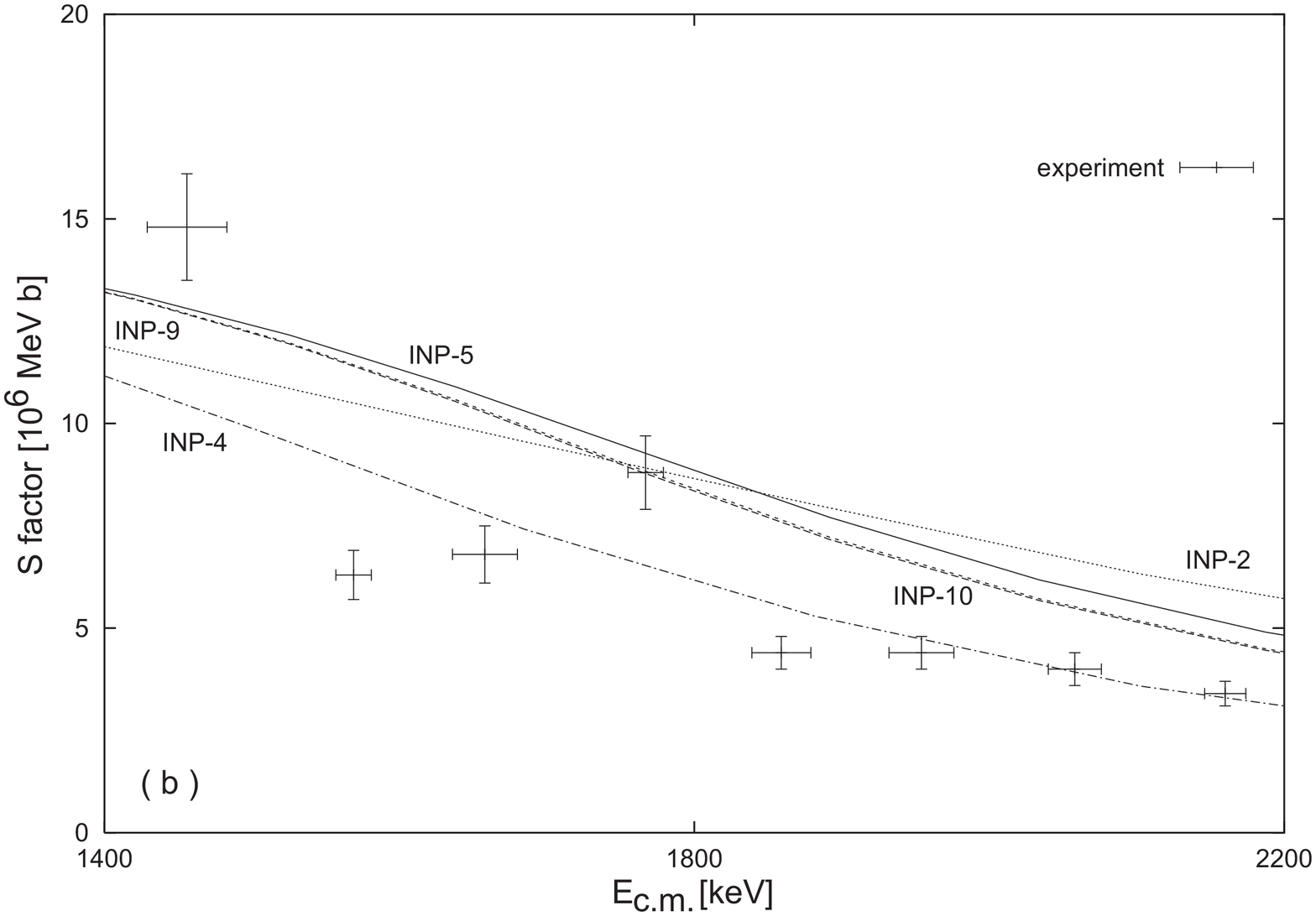}}}
\resizebox{\columnwidth}{!}{\rotatebox{270}{\includegraphics{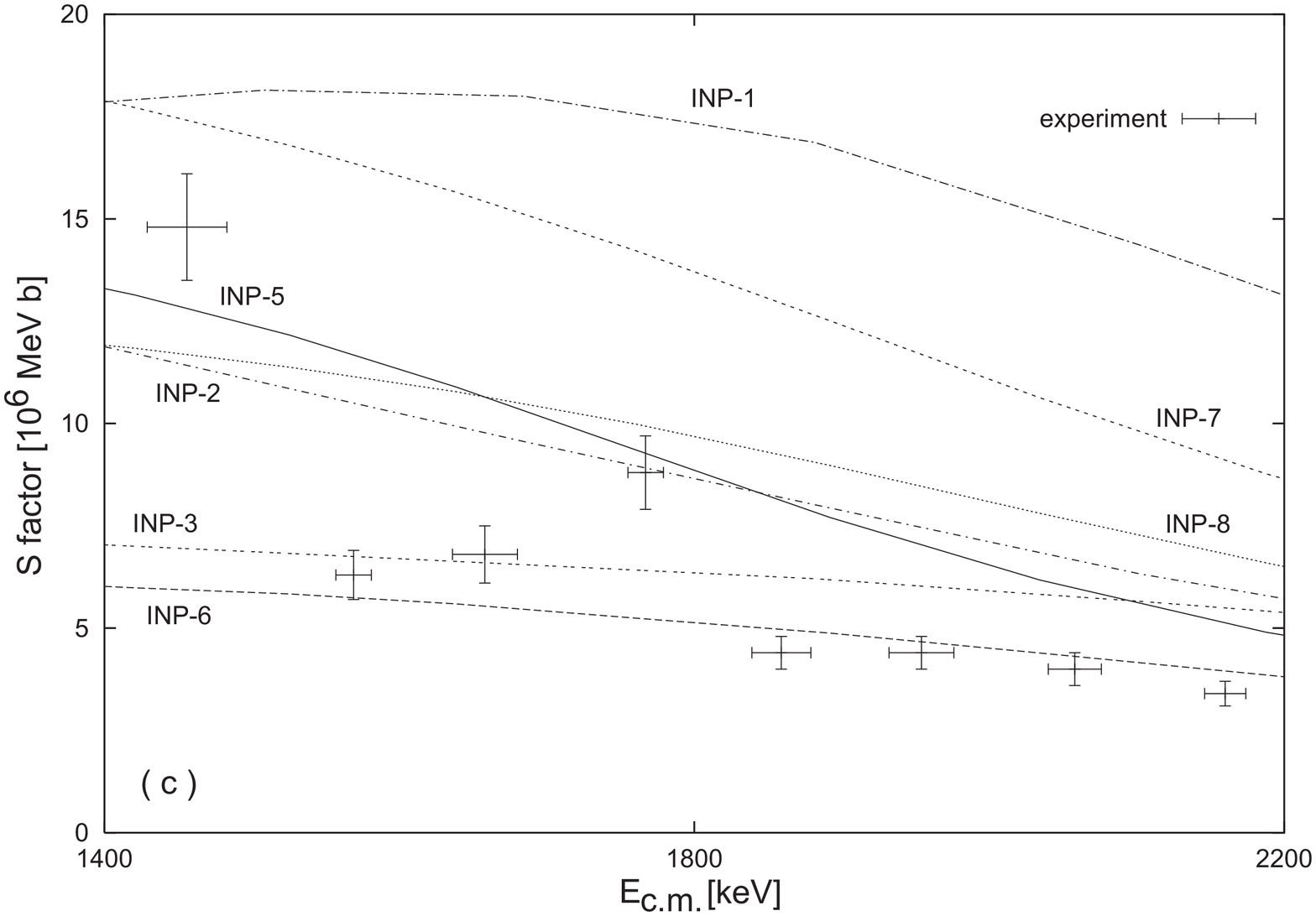}}}
\caption{Experimental $S$ factor of the reaction $^{76}$Se(p,$\gamma$)$^{77}$Br
and the statistical model calculations. The standard predictions of MOST
(INP-1) and NON-SMOKER (INP-5) are compared to the data (a). Below, 
the effect of varying the nuclear level density
with fixed proton optical potential \protect\cite{jeu} is shown (b), 
as well as the effect of varying the optical potential (c). 
In the part (c) of the
figure, the curves labeled INP-1 to INP-3 include the level density of
\protect\cite{dem}, the ones labeled INP-5 to INP-8 the one of
\protect\cite{rau}. See Table \protect\ref{tab:hfinputs} for an explanation
of the labels.}
\label{fig:76}
\end{figure}

The effect of varying the NLDs and the OMPs is shown in
Fig.\ \ref{fig:76}.
The labels are explained in Table \ref{tab:hfinputs}.
What was said for $^{74}$Se(p,$\gamma$) applies equally here. Again,
the dependence on the OMP is the strongest one and a
large spread in absolute values and energy
dependence is found.
The OMPs of \cite{bec} and \cite{bau} significantly overestimate the
data, for the possible reasons mentioned in the
previous section. The OMPs of \cite{holm} and \cite{kon} give a very
flat energy dependence which describes the data well
over most of the energy region. On the other hand, the slightly
steeper excitation functions obtained with the OMPs of
\cite{per} and \cite{jeu} are able to give an overall reasonable account
of the data.

The low-energy structure found in the experiment
(first and fourth data point at the
lowest energies) cannot be reproduced by any of the OMPs and it is doubtful
whether such a behavior of the $S$ factor could be found in a
statistical model calculation at all.

\subsection{$^{82}$Se(p,n)$^{82}$Br}
\label{sec:82}

\begin{figure}
\resizebox{\columnwidth}{!}{\rotatebox{270}{\includegraphics{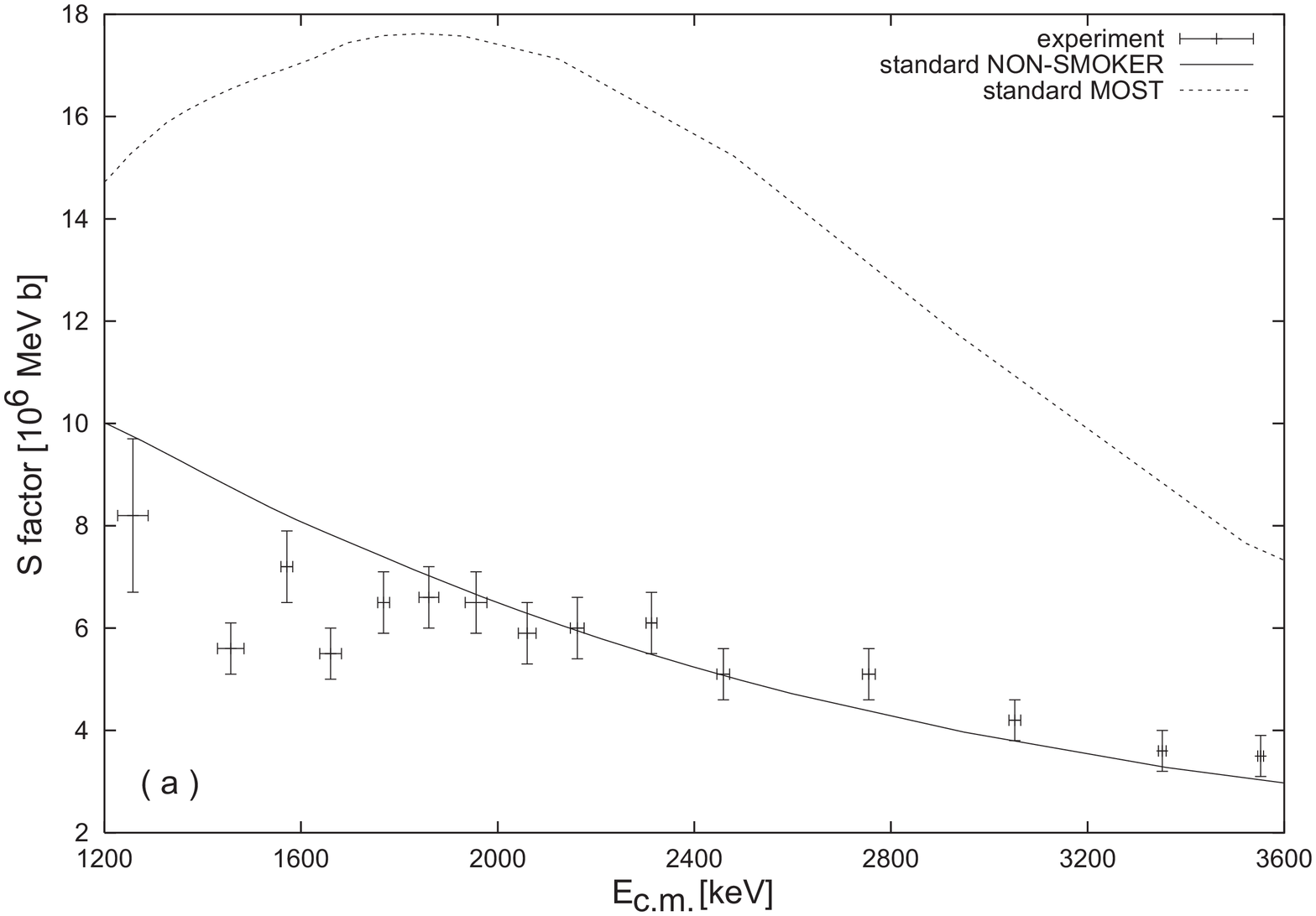}}}
\resizebox{\columnwidth}{!}{\rotatebox{270}{\includegraphics{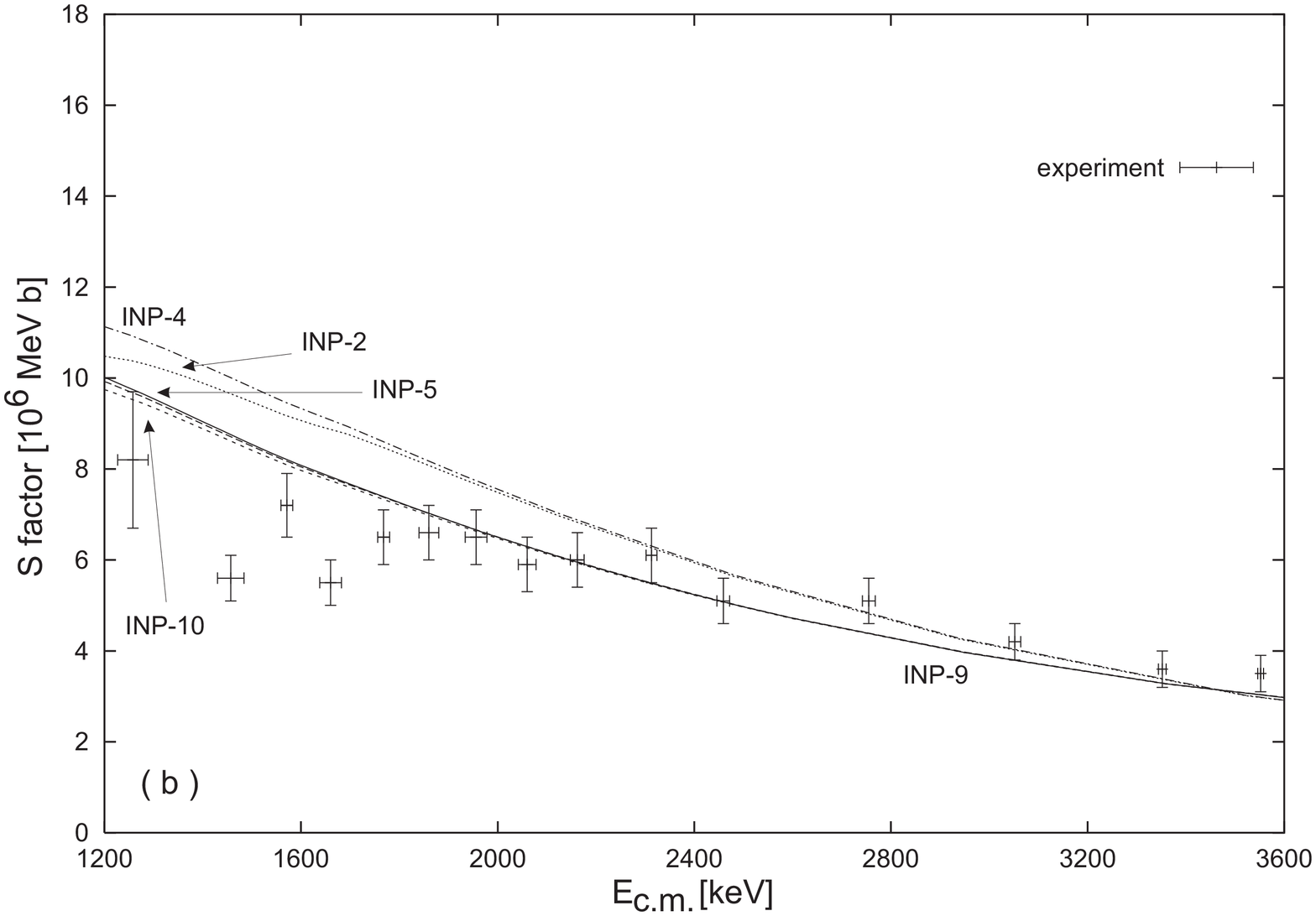}}}
\resizebox{\columnwidth}{!}{\rotatebox{270}{\includegraphics{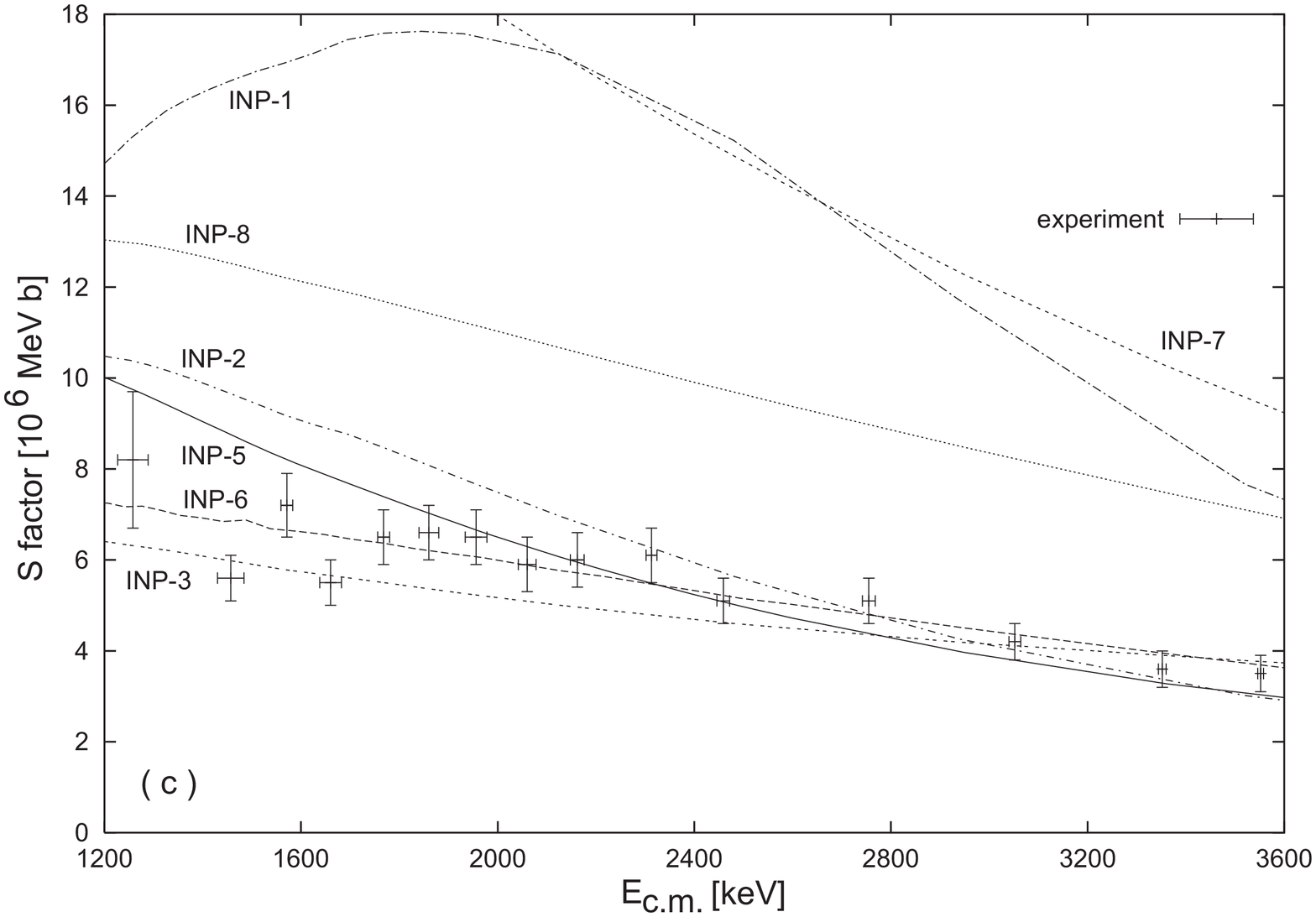}}}
\caption{Experimental $S$ factor of the reaction $^{82}$Se(p,n)$^{82}$Br
and the statistical model calculations. The standard predictions of MOST
(INP-1) and NON-SMOKER (INP-5) are compared to the data (a). Below, 
the effect of varying the nuclear level density
with fixed proton optical potential \protect\cite{jeu} is shown (b), 
as well as the effect of varying the optical potential (c). 
In the part (c) of the
figure, the curves labeled INP-1 to INP-3 include the level density of
\protect\cite{dem}, the ones labeled INP-5 to INP-8 the one of
\protect\cite{rau}. See Table \protect\ref{tab:hfinputs} for an explanation
of the labels.}
\label{fig:82}
\end{figure}

The effect of different NLDs and OMPs is shown in
Fig.\ \ref{fig:82}. The
labels are explained in Table \ref{tab:hfinputs}.
What was said for $^{74}$Se(p,$\gamma$) in Sec.\ \ref{sec:74}
applies similarly here.

For this case, the dependence on the
optical potential
is much stronger than the sensitivity to the NLDs over the whole energy
region. The neutron emission channel opens at the energy of 0.8 MeV and
rapidly becomes the most dominant channel at all measured energies. In
such a case, the HF cross section depends mainly on the incident proton
transmission coefficients which is exactly what is observed in the
figures. Similar to the results obtained for the other Se isotopes, the
OMP of \cite{bau} (INP-1) and \cite{bec} (INP-7)
overpredict the data by at least a factor
of 2. Here, also the potential of \cite{per} (INP-8) yields a significantly
higher $S$ factor although the shape of the energy dependence agrees well.
Much better is the agreement of the OMP of \cite{kon} (INP-3). However, its
energy dependence is slightly too flat. Again, very good agreement
with the data over the whole energy range is found with 
the equivalent square-well potential of \cite{holm} (INP-6).
The microscopic potential of \cite{jeu} (INP-2, INP-5) describes well
the data above 1.7 MeV but shows a different energy dependence below
that energy.

\section{Summary and conclusions}
\label{sec:concl}

We have measured three proton-induced reactions in the astrophysically
relevant energy range:
$^{74,76}$Se(p,$\gamma$)$^{75,77}$Br, and $^{82}$Se(p,n)$^{82}$Br.
Using an activation technique we were able to obtain reaction cross sections
and $S$ factors in the energy range relevant to the astrophysical
$p$ process.
The new data were compared with predictions of the Hauser-Feshbach
statistical theory using global models of OMPs and NLDs. An extensive
investigation of the sensitivity of
the theoretical calculations to the different inputs was presented.
As expected, in this astrophysically relevant low-energy region the results are
more sensitive to the OMP than to the NLD. 
The best overall agreement is obtained with the global microscopic
optical potential of \cite{jeu} and an equivalent square-well potential
\cite{holm}. The latter, somewhat surprising, observation
seems to be in contradiction with a previous
comparison of rates using these two potentials
\cite{hoff}, where the largest deviations between them were found in the mass
range $A>70$. However, this is strongly energy dependent since in Fig.\ 4 of
\cite{hoff} the largest deviations arise at even lower proton energies and the
differences vanish
quickly at higher energies. Likewise, we expect to find larger
differences between the predictions with the two potentials when going
to even lower energy.

Further good agreement with absolute values of the present data
is found with the potential of \cite{kon}. However, the obtained 
energy dependence of the $S$ factor remains slightly too flat.
This is caused by the functional dependence on the energy assumed in the
definition of this potential.

The Se isotopes considered here are deformed and
therefore deformation effects might be important. For the
calculations using the OMPs of \cite{jeu,bec} the well-known approach
of employing a spherical equivalent to a deformed potential (with a
larger diffuseness parameter) was used.
However, from our analyses above it can be seen that even a purely spherical
potential, such as the equivalent square well, can yield satisfactory
results. On the other hand, deformation is important for interpreting
the results obtained with the OMP of \cite{bau}. A significant fraction
of the shown deviations is due to the spherical treatment.
A further investigation of dependencies on the OMP parameters and how to
include deformation effects in OMPs, and more particularly coupled
channels calculations, is beyond the scope of this paper.

Reaction rates are the relevant quantities in astrophysical
applications. In Tables \ref{tab:se74rate}--\ref{tab:se82rate} we give
such astrophysical reaction rates computed from our data including
errors. The temperature range for each reaction was chosen numerically so that 
all significant contributions to the integration over the proton energy
came from within the energy range of our data. Also shown is a
comparison to the rates from the standard NON-SMOKER prediction
\cite{nonsmoker1}; reaction rates for the cases INP-1 to INP-10 can be
derived considering the fact that rates from slowly varying $S$ factors scale
approximately with the $S$ factor. Excellent agreement is found for
$^{74}$Se(p,$\gamma$)$^{75}$Br and $^{82}$Se(p,n)$^{82}$Br. This
illustrates how deviations with respect to the data are averaged
out by the integration involved in the calculation of the reaction
rate, especially at the edge of the respective Gamow window. 
The prediction overestimates the rate of
$^{76}$Se(p,$\gamma$)$^{77}$Br + $^{77}$Se(p,n)$^{77}$Br (Table
\ref{tab:se76rate}) by factors of
1.18--1.77, with better agreement at lower temperature.
Thus, the present work confirms the trend seen in previous investigations of
proton-induced reactions for intermediate mass targets, namely that
there is overall acceptable or good agreement between data and global
predictions.
Apart from a few cases where some deviations seem to persist independent
of nuclear input (Sr isotopes studied in \cite{gyurky01}), the
discrepancies between theoretical calculations are not as large as
those observed for different optical $\alpha$ potentials (see, e.g.,
\cite{som98,gled,dem02}).
This seems to hold for the mass range $70\leq A \leq 100$ 
\cite{ful96,sau97,som98,bork98,chlo99,haris01,gyurky01,ozkan02,galanopulos03}.
Nevertheless,  more measurements are required to completely cover the
relevant mass region and provide constraints on the nuclear input used
in Hauser-Feshbach calculations.

\begin{table}
\caption{Experimental reaction rates computed from the $S$ factor of the
$^{74}$Se(p,$\gamma$)$^{75}$Br reaction and comparison to predicted values.}

\begin{ruledtabular}
\begin{tabular}{rr@{\hspace{0.2cm}$\pm$\hspace{-0.7cm}}lc} 
\multicolumn{1}{c}{\hspace{-0.2cm}$T_9$} &
\multicolumn{2}{c}{\hspace{-0.4cm}Experimental rate} &
\multicolumn{1}{c}{\hspace{-0.5cm}Prediction \protect\cite{nonsmoker1}} \\
\multicolumn{1}{c}{\hspace{-0.2cm}[10$^9$ K]} & \multicolumn{2}{c}{\hspace{-0.4cm}[cm$^3$\,s$^{-1}$\,mol$^{-1}$]} &
\multicolumn{1}{c}{\hspace{-0.5cm}[cm$^3$\,s$^{-1}$\,mol$^{-1}$]} \\
\hline
 2.00& 2.720$\times 10^{1}$&5.301&3.75$\times 10^{1}$\\
 2.50& 2.177$\times 10^{2}$&3.978$\times 10^{1}$&2.55$\times 10^{2}$\\
 3.00& 9.454$\times 10^{2}$&1.663$\times 10^{2}$&1.02$\times 10^{3}$\\
 3.50& 2.813$\times 10^{3}$&4.843$\times 10^{2}$&2.92$\times 10^{3}$\\
 4.00& 6.490$\times 10^{3}$&1.104$\times 10^{3}$&6.75$\times 10^{3}$\\
 4.50& 1.251$\times 10^{4}$&2.115$\times 10^{3}$&1.33$\times 10^{4}$\\
 5.00& 2.117$\times 10^{4}$&3.566$\times 10^{3}$&2.36$\times 10^{4}$\\
 6.00& 4.616$\times 10^{4}$&7.764$\times 10^{3}$&5.79$\times 10^{4}$
\end{tabular} \label{tab:se74rate}
\end{ruledtabular}

\end{table}

\begin{table}
\caption{Experimental reaction rates computed from the $S$ factor of the
reactions $^{76}$Se(p,$\gamma$)$^{77}$Br and
$^{77}$Se(p,n)$^{77}$Br, and comparison to predicted values.}

\begin{ruledtabular}
\begin{tabular}{rr@{\hspace{0.2cm}$\pm$\hspace{-0.7cm}}lc} \multicolumn{1}{c}{\hspace{-0.2cm}$T_9$} &
\multicolumn{2}{c}{\hspace{-0.4cm}Experimental rate} &
 \multicolumn{1}{c}{\hspace{-0.5cm}Prediction \protect\cite{nonsmoker1}}\\
\multicolumn{3}{c}{}&\multicolumn{1}{c}{$\sigma_{\rm (p,\gamma)}+ 0.82
\sigma_{\rm (p,n)}$}\\
\multicolumn{1}{c}{\hspace{-0.2cm}[10$^9$ K]} & \multicolumn{2}{c}{\hspace{-0.4cm}[cm$^3$\,s$^{-1}$\,mol$^{-1}$]} &
\multicolumn{1}{c}{\hspace{-0.5cm}[cm$^3$\,s$^{-1}$\,mol$^{-1}$]} \\
\hline
 2.50& 4.540$\times 10^{2}$&5.671$\times 10^{1}$&6.05$\times 10^{2}$\\
 3.00& 2.189$\times 10^{3}$&2.590$\times 10^{2}$&3.17$\times 10^{3}$\\
 3.50& 7.237$\times 10^{3}$&8.343$\times 10^{2}$&1.14$\times 10^{4}$\\
 4.00& 1.821$\times 10^{4}$&2.078$\times 10^{3}$&3.18$\times 10^{4}$\\
 4.50& 3.758$\times 10^{4}$&4.275$\times 10^{3}$&7.42$\times 10^{4}$
\end{tabular} \label{tab:se76rate}
\end{ruledtabular}

\end{table}

\begin{table}
\caption{Experimental reaction rates computed from the $S$ factor of the
$^{82}$Se(p,n)$^{82}$Br reaction and comparison to predicted values.}

\begin{ruledtabular}
\begin{tabular}{rr@{\hspace{0.2cm}$\pm$\hspace{-0.7cm}}lc} \multicolumn{1}{c}{\hspace{-0.2cm}$T_9$} &
\multicolumn{2}{c}{\hspace{-0.4cm}Experimental rate} &
 \multicolumn{1}{c}{\hspace{-0.5cm}Prediction \protect\cite{nonsmoker1}}\\
\multicolumn{1}{c}{\hspace{-0.2cm}[10$^9$ K]} & \multicolumn{2}{c}{\hspace{-0.4cm}[cm$^3$\,s$^{-1}$\,mol$^{-1}$]} &
\multicolumn{1}{c}{\hspace{-0.5cm}[cm$^3$\,s$^{-1}$\,mol$^{-1}$]} \\
\hline
 1.50& 1.610&1.541$\times 10^{-1}$&2.07$\times 10^{0}$\\
 2.00& 4.561$\times 10^{1}$&4.639&4.91$\times 10^{1}$\\
 2.50& 4.377$\times 10^{2}$&4.515$\times 10^{1}$&4.45$\times 10^{2}$\\
 3.00& 2.231$\times 10^{3}$&2.333$\times 10^{2}$&2.32$\times 10^{3}$\\
 3.50& 7.519$\times 10^{3}$&7.982$\times 10^{2}$&8.52$\times 10^{3}$\\
 4.00& 1.907$\times 10^{4}$&2.054$\times 10^{3}$&2.45$\times 10^{4}$\\
\end{tabular} \label{tab:se82rate}
\end{ruledtabular}

\end{table}

\begin{acknowledgments}
This work was partially supported by OTKA (T034259, T042733, F043408),
the NATO CRG program (CRG961086), the Greek-Hungarian bilateral
collaboration program (GSRT/Demokritos/E797),
and the Swiss NSF (2000-061031.02). 
The authors are grateful to 
I.~Borb\'ely-Kiss
for carrying out the PIXE measurements. Gy.\ Gy.\ 
and Zs.\ F. are Bolyai fellows. T. R. is a PROFIL professor (Swiss NSF
2024-067428.01).
\end{acknowledgments}

\end{document}